\documentclass[a4paper,english,aps,prb,twocolumn,floatfix,showpacs,amsfonts,amssymb,superscriptaddress]{revtex4} 
\usepackage[T1]{fontenc}
\usepackage[latin1]{inputenc}
\usepackage{graphics}
\usepackage{amssymb}
\usepackage{amsmath}
\usepackage{overpic}

\newcommand{\be}{\begin{equation}}
\newcommand{\ee}{\end{equation}}
\newcommand{\bea}{\begin{eqnarray}}
\newcommand{\eea}{\end{eqnarray}}

\def\a{\alpha}

\def\e{\varepsilon}
\def\d{\delta}

\def\m{\mu}
\def\l{\lambda}

\def\o{\omega}

\def\s{\sigma}

\def\L{\Lambda}

\def\ra{\rightarrow}

\def\nn{\nonumber}
\def\lb{\label}
\def\pref#1{(\ref{#1})}

\newcount\bozza \bozza=1
\ifnum\bozza=0
\newdimen\shift \shift=-2truecm
\def\lb#1{%
{\label{#1}\rlap{\kern\shift{$\scriptstyle#1$}}}}
\else\def\lb#1{\label{#1}} \fi

\begin{document}

\title{Slowing down of vortex motion in thin NbN films \\near the 
  superconductor-insulator transition}

\author{Rini Ganguly}
\affiliation{Department of Condensed Matter Physics and Materials Science, Tata Institute of Fundamental Research,
Homi Bhabha Rd., Colaba, Mumbai 400 005, India}

\author{Dipanjan Chaudhuri}
\altaffiliation{Permanent affiliation: Indian Institute of Science Education And Research-Kolkata, Mohanpur - 741 246, West Bengal, India}
\affiliation{Department of Condensed Matter Physics and Materials Science, Tata Institute of Fundamental Research,
Homi Bhabha Rd., Colaba, Mumbai 400 005, India}

\author{Pratap Raychaudhuri}
\affiliation{Department of Condensed Matter Physics and Materials Science, Tata Institute of Fundamental Research,
Homi Bhabha Rd., Colaba, Mumbai 400 005, India}

\author{Lara Benfatto}
\affiliation
{CNR-ISC and Department of Physics, Sapienza University of Rome, P. le Aldo Moro 5, 00185, Rome, Italy}


\date{\today}

\begin{abstract}
  We present a quantitative comparison between the measurements of the
  complex conductance at low (kHz) and high (GHz) frequency in a thin
  superconducting film of NbN and the theoretical predictions of the
  dynamical Beresinksii-Kosterlitz-Thouless theory. While the data in the
  GHz regime can be well reproduced by extending the standard approach to
  the realistic case of a inhomogeneous sample, the low-frequency measurements
  present an anomalously large dissipative response around $T_c$. This
  anomaly can only be accounted for by assuming a strong slowing down of
  the vortex diffusion in the kHz regime, or analogously a strong
  reduction of the length scale probed by the incoming finite-frequency
  field. This effect suggests the emergence of an intrinsic length scale
  for the vortex motion that coincides with the typical size of
  inhomogeneity probed by STM measurements in disordered NbN films.

\end{abstract}


\maketitle

\section{Introduction}

More than forty years after its discovery,\cite{review_minnaghen,jose_book} the
Beresinksii-Kosterlitz-Thouless (BKT) transition\cite{bkt,bkt2} still
represents one of the most fascinating phenomena in condensed-matter
systems. Indeed, it describes the universality class for the phase
transition in two spatial dimensions in a system displaying $U(1)$
symmetry. After its original formulation within the classical $XY$ spin
model, it has been later on applied to a wide class of phenomena, mostly
related to the superfluid or superconducting (SC) transition in two
dimensions, as it occurs in artificial Josephson junctions, thin films, and
recently also cold atoms.\cite{jose_book} In all these cases the transition
is driven by topological vortex excitations instead of the usual vanishing
of the order parameter, leading to striking predictions for the behaviour
of several physical quantities.

The most famous hallmark of BKT physics is certainly the discontinuous but
universal jump of the density of superfluid carriers at the
transition,\cite{nelson_prl77} that has been successfully observed in
superfluid He films.\cite{helium4} However in the case of superconducting
materials the superfluid-density jump turned out to be rather elusive:
indeed, while some signatures have been identified in thin films of
conventional superconductors as MoGe\cite{MoGe,yazdani_prl13},
InOx\cite{fiory_prb83,armitage_prb07,armitage_prb11} and
NbN,\cite{kamlapure_apl10,mondal_bkt_prl11,yong_prb13} they are less
evident in other cases as thin films of high-temperature cuprate
superconductors\cite{lemberger_prb12} or SC interfaces between
oxides.\cite{moler_prb12} The lack of clear BKT signatures is due in part
to two intrinsic characteristics of the SC films, absent in superfluid
ones:\cite{review_minnaghen} (i) the presence of quasiparticle excitations,
that contribute to the decrease of the superfluid density limiting the
observation of BKT effects to a small temperature range between $T_{BKT}$
and the BCS critical temperature $T_c$, and (ii) the screening effects
due to charged supercurrents. The latter can be avoided by decreasing the
film thickness $d$, so that the Pearl length\cite{review_minnaghen} $\L=2\l^2/d$, where $\l$ is
the penetration depth, exceeds the sample size making the interaction
between vortices effectively long-ranged.  However, this also implies that
in the case of conventional superconductors, like InOx and NbN, the
observation of BKT physics is restricted to samples near to the
superconductor-to-insulator transition (SIT), where several additional
features must be considered along with the presence of BKT vortices. The
most important one is the emergence at strong disorder of an intrinsic
inhomogeneity of the SC properties, as shown in the last few years by
detailed tunneling spectroscopy
measurements.\cite{sacepe_11,mondal_prl11,pratap_13,noat_prb13}

Such intrinsic granularity of disordered SC films, that has been
interpreted theoretically as the compromise between charge localization and
pair hopping near the
SIT,\cite{ioffe,nandini_natphys11,seibold_prl12,lemarie_prb13} has been
efficiently incorporated in the BKT description of the superfluid density
as an average of the superfluid response over the distribution of local
critical temperatures.\cite{benfatto_prb08,review13} This leads to a drastic smearing of
the superfluid-density jump predicted by the conventional BKT theory, in
accordance with the experimental observations of the inductive response in
thin films of conventional
superconductors,\cite{mondal_bkt_prl11,yong_prb13} as measured by means of
two-coil experiments. Notice that this experimental set-up measures the
complex conductivity $\s(\o)$ of a superconductor at low but finite
frequency $\o$, usually $\o\simeq 50-60$ kHz. As a consequence, along with
the superfluid response, connected to the imaginary part $\s_2(\o)$, it
also allows one to measure the dissipative part $\s_1(\o)$, that displays a
peak slightly above $T_{BKT}$, whose width in temperature correlates
usually with the broadening of the superfluid-density jump.\cite{yong_prb13}

According to the standard view\cite{ambegaokar_prb80,HN} the largest
contribution to $\s_1(\o)$ near $T_{BKT}$ is expected to come from the same
vortex excitations that control the suppression of the superfluid density.
Indeed, the finite dissipation comes essentially from the cores of the
vortices thermally excited above $T_{BKT}$, that can move at finite probing
frequency over a length scale $r_\o$ of the order of $r_\o\simeq
\sqrt{D/\o}$. Here $D$ is the vortex diffusion constant, that is usually
assumed\cite{ambegaokar_prb80} to coincide with the electron diffusion
constant. The maximum of $\sigma_1$ is then expected to occur at the
temperature $\bar T$ above $T_{BKT}$ where $\xi(\bar T)\simeq r_\omega$,
where $\xi(T)$ is the BKT correlation length. As a consequence, one would
expect a shift or the $\sigma_1$ maximum towards higher temperatures, along with a broadening 
and enhancement of the dissipative peak as $\o$ increases. As we will show
in this paper, these conditions are strongly violated in thin films of
disordered NbN.  By comparing the experimental results obtained on the same
sample measured both at low (10 to 100 kHz) and high (1 to 10 GHz)
frequency we show that the dissipative response is approximately the same
in the two regimes. When compared with the theoretical predictions for the
BKT transition in an inhomogeneous system these results imply that the
dissipation observed in the kHz regime is anomalously large, as reported
before also MoGe and InO films,\cite{MoGe} while in the GHz regime the
observed resistive contribution agrees approximately with the BKT
expectation. Within our approach the large resistive contribution observed
in the kHz range can be accounted for only by assuming a strong reduction
of the vortex difussion constant with respect to the conventional value of
the Bardeen-Stephen theory. Such a slowing down of vortices at low
frequency can be also rephrased as the emergence of an {\em intrinsic}
length scale cut-off $\xi_V$ for vortex diffusion in our disordered thin
films, that makes the dissipative response quantitatively very similar in
the two range of frequencies. More importantly, $\xi_V$ correlates well
with the typical size of SC islands observed in similar samples by STM,\cite{mondal_prl11,pratap_13}
showing that the disorder-induced inhomogeneity is a crucial and
unavoidable ingredient to understand the occurrence of BKT physics in thin
films.

The plan of the paper is the following. In Sec. II we present the
experimental results obtained by means of two-coils mutual inductance
technique in the kHz and by means of measurements in the Corbino geometry
in the microwave. The direct comparison between the experimental data
obtained on the same sample in the two regimes of frequencies shows clearly
the emergence of an anomalously large dissipative response at low
frequency. To quantify this anomaly we introduce in Sec. III the standard
BKT approach for the finite-frequency response in a homogeneous system, and
show its failure to reproduce the experimental data. In Sec. III we extend
the finite-frequency BKT approach to include the effect of inhomogeneity,
along the line of previous work done in the static
case.\cite{mondal_bkt_prl11,review13,yong_prb13}. Within this scheme we
discuss the role played by the anomalous vortex diffusion constant at low
frequency, and we comment on its relation to real-space structures due to
disorder. The final remarks are presented in Sec. V. Finally, Appendix A
contains some additional technical detail on the description of the BKT
physics at finite frequency.

\section{Experimental results}
\subsection{Details of the measurements and analysis}

The electrodynamic response was studied on a 3 nm thick NbN sample, that we expect to be in the 2D limit as previous measurements in analogous samples have shown.\cite{mondal_bkt_prl11} The superconducting
BKT transition was studied in both kHz(10 to 100 kHz) and microwave(1 GHz
to 10 GHz) frequency range on the same sample to explore vortex diffusion
as a function of the probing length scale.

The kHz data were acquired using a home-built two-coil mutual inductance
set-up,\cite{kamlapure_apl10} where we drive the primary coil at a desired
frequency varying from 10-100 kHz. The amplitude of ac excitation is kept
at 10 mOe. The sample here is a circle of diameter of 8 mm, which we
prepare by DC magnetron sputtering of Nb on MgO(100) in an argon-nitrogen
gas mixture. We place the sample in between the coaxial primary and
secondary coil, and we measure the induced voltage of the secondary
coil. Since the degree of coupling between the coils varies with
temperature due to the variation of complex penetration depth $\l_\o$ ( $\l_\o=
(\l^{-2} + i \d^{-2})^{-1/2}$,  see Eqs.\ \pref{lambda}-\pref{delta} below)
of the superconducting film, the real and imaginary part of the voltage
induced in the secondary coil gives the complex mutual inductance $M_{exp}$
of the coils as a function of temperature. The theoretical value of
$M_{theo}$ as a function of $\l_w$ can be determined by solving numerically
the coupled set of Maxwell and London equations for the particular coil and
sample geometry of our set-up. The numerical method takes into account the
effect of the finite radius of the film, as proposed by
J. Turneaure,\cite{Turneaure1,Turneaure2} see also Ref.\
[\onlinecite{mintuthesis}]. We obtain a 2D matrix (typically 100 $\times$
100) of complex mutual inductance values for different sets of Re($\l_w ^{-2}$)
and Im($\l_w ^{-2}$). Then we compare $M_{exp}$ with the calculated $M_{theo}$
in order to extract $\l_w$ as a function of temperature.
  
Microwave spectroscopy was carried out in Corbino
geometry\cite{mondal_scire13} on the same piece of sample after cutting it
in 5 mm $\times$ 5 mm size and thermally evaporating Ag on it. By using the
same piece of the sample we can avoid any effect of change in the SC
properties (SC gap, superfluid stiffness, etc.)  while studying two
different frequency regimes. In this way we can attribute the change of the
optical response only to variations of the vortex diffusion constant, which
is the aim of the present work. The sample here terminates a 1 m long ss
coaxial cable to reflect the microwave signal, generated internally from a
vector network analyser (VNA) spanning 10 MHz to 20 GHz.  The complex
reflection coefficient($S_{11}$) measured by the VNA is first corrected
using three error coefficients for the cable, which we get after
calibration with three standards.\cite{ohashi,dressel,booth} To calibrate
the cable at experimental condition, i.e at low temperature, we use as
short standard the spectrum of a thick ordered NbN sample taken at the
lowest temperature, and as loads the sample spectra taken at two different
temperatures above $T_c$. Such a calibration technique is less prone to
error, since two of the three calibrators are measured during the same
thermal cycle with the actual sample. The corrected $S_{11}$ is then related to
the complex impedance $Z$ of the sample by means of the relation:
\begin{equation}
S_{11} = \frac{(Z - Z_0)}{(Z + Z_0)}
\end{equation}
where $Z_0 $ is the characteristic impedance of the cable, i.e. 50
$\Omega$ in our case. The complex conductivity $\s$ of the sample is the given
by
\begin{equation}
\s = \frac{\ln(b/a)}{2 \pi d Z}
\end{equation}
where $a$ and $b$ are the inner and outer radius of the film, and $d$ is the thickness.

\subsection{Experimental data}

\begin{figure}[htp]
\includegraphics[scale=0.33,angle=0,clip=]{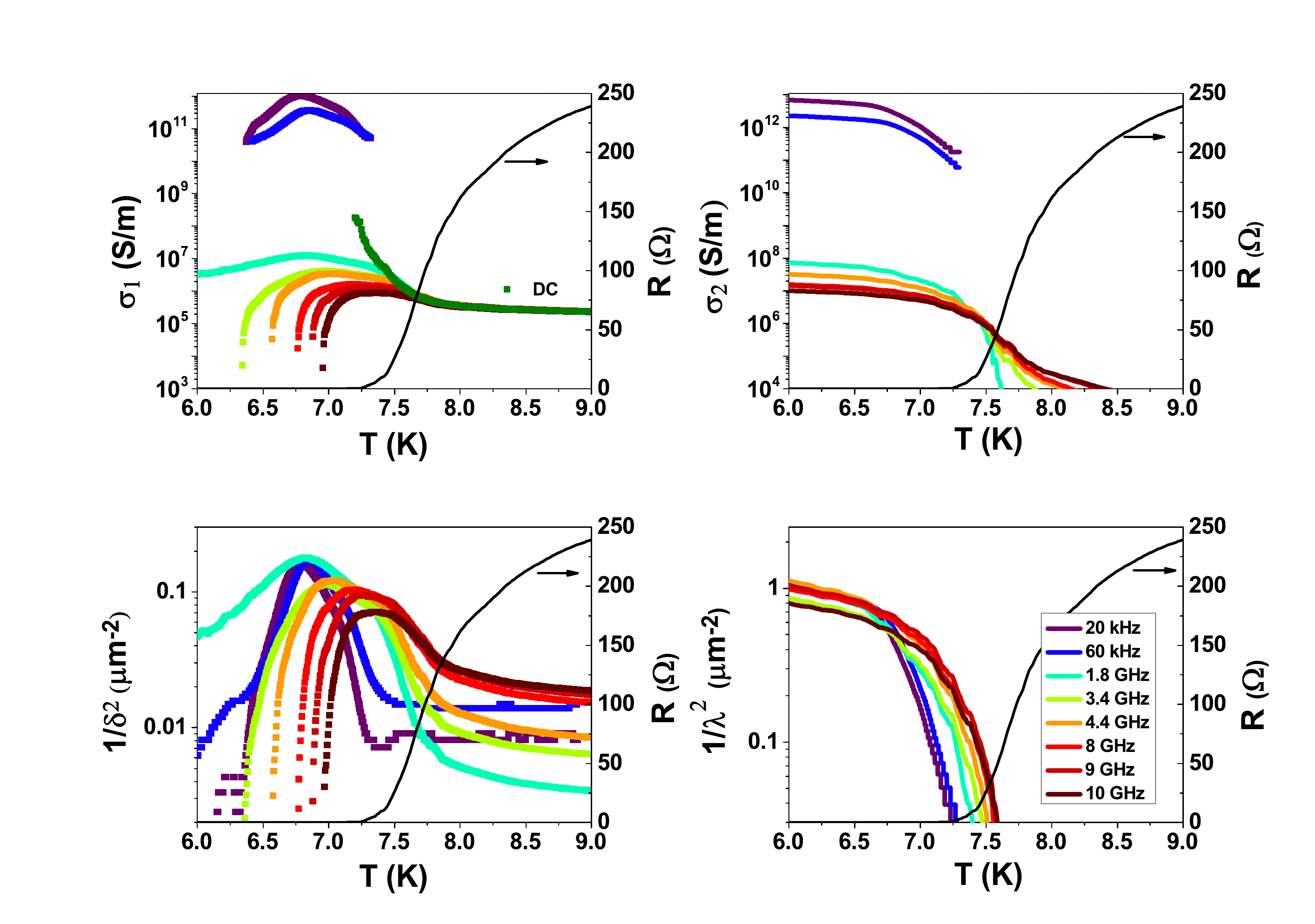}
\caption{(Color online) Experimental data of $\s_1$ and $\s_2$ at both kHz
  and GHz frequency range(panel (a) and (b)).Panel (c) and (d) show the
  comparison of $\d^{-2}$ and $\l^{-2}$ at similar frequencies. The panels
  contain also the result of simultaneous dc resistance measurement. Notice that $\s_1$ in the microwave regime matches perfectly above $T_c$ the value of the dc conductivity obtained by the measured resistance, see panel (a).}
\label{fig-expt}
\end{figure}

\begin{figure}[htp]
\includegraphics[scale=0.33,angle=0,clip=]{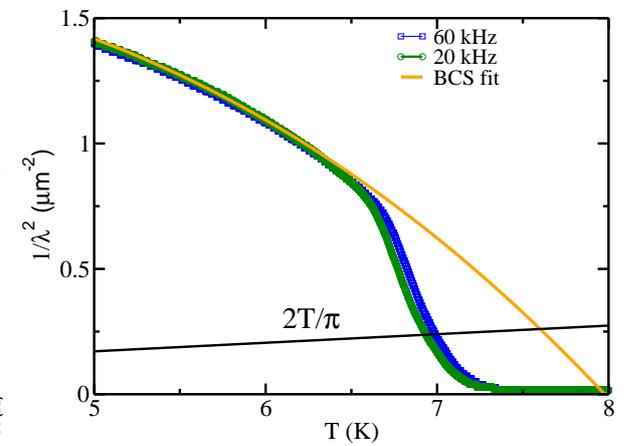}
\caption{(Color online) Inverse penetration depth at low frequency compared
  with a BCS fit of the low-temperature part. The deviation from the BCS
  fit, which occurs slightly before than the intersection with the 
  universal line $2T/\pi$, signals the occurrence of a BKT transition due
  to thermally excited vortices. The ratio $\Delta(0)/T_c=2.2$ obtained
  from the fit, with $T_c$ BCS
  temperature, agrees well with the estimate given by tunnelling experiments
  in thick NbN films.\cite{mondal_prl11}}
\label{fig-expt2}
\end{figure}

The results of the kHz and GHz measurement of the complex conductivity are
shown in Fig.\ \ref{fig-expt}a,b. Notice that even though the kHz
measurements lose sensitivity away from $T_c$, our microwave measurements
can capture the normal state conductivity quite well, and it exactly matches  the one obtained 
from dc measurement (see Fig.\ \ref{fig-expt}a). To compare data at
different frequencies it is more convenient to convert the complex
conductivity in a length scale:
\bea
\lb{eql}
\l^{-2}&=&\m_0 \omega\sigma_2,\\
\lb{eqd}
\delta^{-2}&=&\m_0 \omega\sigma_1,
\eea
where $\l^{-2}$ coincides with the usual SC penetration depth, proportional
to the superfluid density of the sample (see Eq.\ \pref{defjs} below). At
finite frequency $\l^{-2}$ persists slightly above $T_c$, in a temperature
range that increases proportionally to the probing frequency, as observed
also in thick samples,\cite{mondal_scire13} and as expected by scaling
near criticality.\cite{ffh} This effect, shown in Fig.\ \ref{fig-expt}d, is
negligible for low frequency and becomes appreciable in the microwave
regime. The low-temperature part of $\l^{-2}$ can be fitted very well by
means of a BCS formula, as shown in Fig.\ \ref{fig-expt2}. However, near
$T_c$ a sudden deviation of $\l^{-2}$ from the BCS fit occurs, signaling
the occurrence of a vortex-induced BKT transition. As already observed
before,\cite{mondal_bkt_prl11,yong_prb13} the low vortex fugacity of NbN
films moves the BKT transition at temperatures slightly smaller than the
one where the BCS curve intersects the universal BKT line (see also Eq.\
\pref{defjs} below).

The length $\delta^{-2}$ is instead a
measure of the fluctuations around $T_c$, that originate in our BKT sample
by the vortex motion at finite frequency, as we shall discuss
below. On very general ground, one can associate the probing frequency $\o$
with a finite length scale $r_\o$ by means of the diffusion coefficient $D$:\cite{ambegaokar_prb80}
\be
\lb{defro}
r_\o=\sqrt{\frac{14D}{\o}}.
\ee
Within the standard Bardeen-Stephen model\cite{HN} the vortex motion causes
dissipation because of the normal-electron component present in the vortex
cores. Thus, the diffusion constant $D$ in Eq.\ \pref{defro} should scale
with the electron diffusion constant, that can be estimated from the Fermi
velocity and the mean free path obtained by resistivity and Hall
measurements at 285K, as $D\simeq 10^{-5} m^2/s$. Thus, in the kHz regime
$r_\o$ should approch the system size, giving negligible dissipative
effects, in sharp contrast to the experimental observation.  Indeed, the intensity of $\d^{-2}$ and the peak width are similar
both in the kHz and the GHz regime, despite a change of frequency by six
orders of magnitude, see Fig.\ \ref{fig-expt}d and\ \ref{fig-expt}c).
 As we shall discuss in the next section, the anomalously
large dissipative response found in the kHz frequency regime is the
hallmark that inhomogeneity cut-off the vortex diffusion at scales $r_\o\sim \xi_V$ of order
of the typical size $\xi_{inh}$ of the inhomogeneous domains. Thus, while
in the GHz regime a standard value of the diffusion constant leads to a
probing length $r_\o$ that is already of the order of $\xi_{inh}$, leading
to a dissipative response in good quantitative agreement with standard
predictions for the BKT theory in an inhomogeneous system, in the kHz regime
the same approach fails, unless one assumes a diffusion constant $D$ much
smaller than what predicted by the Bardeen-Stephen theory.

\section{Dynamical BKT theory: the conventional view}

\begin{figure}[htp]
\includegraphics[scale=0.3,angle=0,clip=]{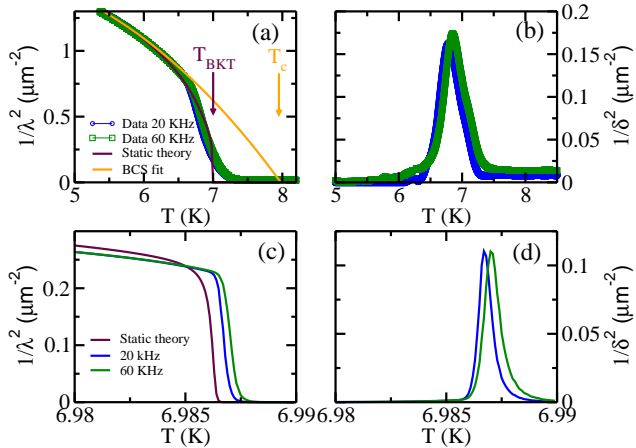}
\caption{(Color online) Comparison between the experimental measurements
  (symbols) of
  $\l^{-2}$ (a) and $\d^{-2}$ (b) and the predictions (solid lines) (c,d) of the 
  conventional dynamical BKT theory for $D=10^{-5}$m$^2/$s.The BCS fit in (a)  works well far from the
  transition and allows one to estimate the BCS transition temperature
  $T_c$, while $T_{BKT}$ (obtained by using $\mu/J=1.2$) marks here the prediction of the homogeneous,
  static case. By using a conventional estimate of the vortex diffusion
  constant one obtains the rounding of $\l^{-2}$ shown in panel (c): as one
  can see it occurs on a temperature scale two orders of magnitude smaller than
  the one observed experimentally. At the same time, the peak in $\d^{-2}$
  shown in panel (d) is essentially a delta-like peak when compared to the
  measurements reported in panel (b).}
\label{fig-stand}
\end{figure}

The extension of the BKT theory to include dynamics effects was developed
soon after its discovery in a couple of seminal papers by Ambegaokar et
al., \cite{ambegaokar_prb80} who considered the case of superfluid Helium,
and afterwards by Halperin and Nelson,\cite{HN} who extended it to charged
superconductors. The effect of the transverse motion of vortices
under an applied electric field is then encoded in an effective
frequency-dependence dielectric function $\e(\o)$, in analogy with the
motion of the charges for the Coulomb plasma. The resulting complex
conductivity of the film can be expressed as:\cite{HN,review_minnaghen}
\be
\lb{defs}
\s(\o)=-\frac{n^0_s d e^2}{i\o m \e(\o)},
\ee
$m$ the electron mass and $n_s^0$ is the mean-field superfluid density, i.e. the one including 
BCS quasiparticle excitations but not the effect of vortices. The dielectric function
is controlled by the fundamental scaling variables appearing in the BKT
theory,\cite{review_minnaghen,review13} i.e. the bare superfluid stiffness $J$ and the vortex fugacity $g$, defined as usual by
\be
\lb{defj}
J=\frac{\hbar^2 n^0_s d}{4m}, \quad g=2\pi
e^{-\beta \mu},
\ee
where $\mu$ is the vortex-core energy.  As it is well
known,\cite{bkt2,review_minnaghen,review13} the role of vortices at large distances can be fully captured
  by the renormalization-group (RG) equations of the BKT theory for the two
  quantities $g$ and $K\equiv \pi J/T$:
\bea
\lb{eqk} 
\frac{dK}{d\ell}&=&-K^2g^2,\\ 
\lb{eqg} 
\frac{dg}{d\ell}&=&(2-K)g,
\eea
where $\ell=\ln (a/\xi_0)$ is the RG-scaled lattice spacing with respect to
the coherence length $\xi_0$, that controls the vortex sizes and appears as
a short-scale cut-off for the theory. From Eq.\ \pref{defs} we can derive
the real and imaginary part of the conductivity in terms of the bare
stiffness $J$ and the dielectric function, so that one has, in agreement
with Eqs.\ \pref{eql}-\pref{eqd} above:
\bea
\lb{lambda}
\l^{-2}&=&\m_0\o\sigma_2=\frac{J}{d\a} \mathrm {Re}\frac{1}{\e}=\frac{J}{d\a}\frac{\e_1}{\e_1^2+\e_2^2},\\
\lb{delta}
\d^{-2}&=&\m_0\o\sigma_1=-\frac{J}{d\a} \mathrm {Im}\frac{1}{\e}=\frac{J}{d\a}\frac{\e_2}{\e_1^2+\e_2^2}.
\eea
where $\alpha=\hbar^2/4e^2\m_0$ is a
numerical factor. In particular, if $\l$ is expressed in $\m$m, $d$ in \AA \ and $J$ in K, then $\a=0.62$.
As we discuss in details in Appendix A, $\e(\o)$  is a
function of both $K$ and $g$. In particular, in the static limit one can
show that $\e$ is purely real, and it is given by:
\be
\lb{eps0}
\e(\o=0)=\e_1=\frac{K(0)}{K(\ell\ra\infty)}=\frac{J(0)}{J(\ell\ra\infty)},
\ee
so that $\d^{-2}=0$ and the inverse penetration depth $\l^{-2}$ is
controlled by the renormalized stiffness $J_s$ introduced in
Refs. [\onlinecite{nelson_prl77,review13}]:
\be
\lb{defjs}
J_s\equiv \frac{T K(\ell=\infty)}{\pi}\equiv \frac{\hbar^2 n_s d}{4m}=\frac{\a d}{\l^2}.
\ee
with $n_s$ real superfluid density, including also vortex-excitation effects.
According to the RG equations \pref{eqk}, when $K\gtrsim 2$ the fugacity
$g$ flows to zero at large distances, so that $K(\ell\ra\infty)\neq 0$. The
resulting  $J_s$ is finite but in general smaller than its BCS counterpart $J$,
due to the effect of bound vortex-antivortex pairs at short length scales, as it
has been discussed in the context of NbN thin
films.\cite{mondal_bkt_prl11,yong_prb13} Instead when $K\lesssim 2$, $g$
diverges, signalling the proliferation of free vortices. The BKT transition temperature is the one where $K(\ell=\infty)=2$, so that at $T_{BKT}$ $J_s$ if finite and it jumps discontinuously to zero right above it:
\be
\lb{jump}
J_s(T_{BKT}^-)=\frac{2T_{BKT}}{2}, \quad J_s(T_{BKT}^+)=0.
\ee
At finite frequency $\e(\o)$ develops an imaginary part due to the vortex motion: in
first approximation (see also Appendix A) one can put\cite{ambegaokar_prb80}
\be
\lb{e2app}
\e_2\simeq \frac{Dn_f}{\o}=\frac{D}{\o \xi^2} \sim \left(\frac{r_\o}{\xi}\right)^2,
\ee
where $n_f$ is the free-vortex density, expressed in terms of the vortex
correlation length $\xi$, and $r_\o$ is the frequency-dependent probing
length scale introduce in Eq.\ \pref{defro} above. The length scale $r_\o$ provides also a cut-off for the real part of the dielectric function, that is given approximately by Eq.\
\pref{eps0} with $\ell=\infty$ replaced by $\ell_\o=\ln (r_\o/\xi_0)$:
\be
\e_1(\o)\simeq \frac{J(0)}{J(\ell_\o)},
\ee
so that instead of the discontinuous
divergence of $\e_1$ expected for $\o=0$, due to the superfluid-density jump \pref{jump}, one finds a rapid increase across
$T_{BKT}$. The resulting temperature dependence of $\l^{-2}$ and $\d^{-2}$ in Eqs.\ \pref{lambda}-\pref{delta} is controlled by the increase of $\e_1,\e_2$ across $T_{BKT}$. In particular, since $\e_2$ from Eq.\ \pref{e2app} becomes sizeable
when one moves away from $T_{BKT}$ due to the proliferation of free
vortices, until it overcomes $\e_1$, $\l^{-2}$
displays a rapid downturn instead of the
discontinuous jump \pref{jump} of the static theory. Instead $\d^{-2}$ in Eq.\
\pref{delta} starts to increase at $T_{BKT}$ and shows a maximum at approximately the temperature $\bar T$  where 
$\e_2\simeq \e_1\sim
{\cal O}(1)$.  In terms of the characteristic length scales appearing in
Eq.\ \pref{e2app} this occurs when
\be
\lb{tbar}
\xi(\bar T)\approx r_\o.
\ee
The correlation length within the BKT theory is described by an
exponentially-activated behavior\cite{bkt,review_minnaghen,HN,review13} 
\be
\lb{xi}
\xi\simeq A \xi_0 \exp\left(\frac{b}{\sqrt{t}}\right),
\ee
where the coefficient $b$ is connected to the distance between the
$T_{BKT}$ and the mean-field temperature $T_c$, and to the vortex-core
energy:\cite{HN,benfatto_prb09}
\be
b\simeq \frac{4
}{\pi^2}\frac{\mu}{J}\sqrt{t_c}, \quad t_c={\frac{T_c-T_{BKT}}{T_{BKT}}}.
\ee
By means of Eq.\ \pref{xi}, and using $\mu/J\simeq 1$ as evidenced by
the analysis of the $\l^{-2}$ far from the transition regime we are
investigating,\cite{mondal_prl11} we then obtain that up to multiplicative factors of
order one the transition width at finite frequency is approximately:
\be
\lb{dtbar}
\frac{\bar T-T_{BKT}}{T_{BKT}}\simeq \frac{1}{[\ln (r_\o/\xi_0)]^2} t_c.
\ee
The above Eq.\ \pref{dtbar} confirms the general expectation that
the broadening of the transition due
to finite-frequency effects is fully controlled by the probing length scale
$r_\o$, where the diffusion constant enters in a crucial way, see Eq.\
\pref{defro}. Let us then start with an estimate of the finite-frequency
effects in NbN based on the standard value of the diffusion constant
given by the Bardeen-Stephen model, where $D$ coincides
with the diffusion constant of electrons $D_e\simeq v_F
\ell_{loc}$, that is around $D_e\simeq 10^{-5}m^2/s$ in
NbN.\cite{mondal_scire13} The distance between $T_c$ and $T_{BKT}$ can be
estimated by a BCS fit of the data at low temperatures, as shown in Fig.\
\ref{fig-stand}a, and it is in the present case $t_c=0.1$. Finally, for $\xi_0\simeq 10$ nm, as appropriate for
NbN,\cite{xi0} one obtains from Eq.\ \pref{tbar} 
\be
\lb{est}
\bar T-T_{BKT}\simeq 0.01 K.
\ee
This estimate is confirmed by the numerical calculation of the optical response based on the full 
expression of the dielectric function $\e(\o)$ reported in Appendix A,
and shown in Fig.\ \ref{fig-stand}c,d. As one can see, finite-frequency
effects lead indeed to a neglegible smoothening of the superfluid-density jump with
respect to the static case, and to a finite dissipative response $\d^{-2}$ whose width in temperature 
is two order of magnitude {\em smaller} than what observed in
real data, reported in Fig.\ \ref{fig-stand}b. The result for $\l^{-2}$ can be easily understood by comparing the scale $r_\o$  with the other two length scales that act as cut-off on the RG equations \pref{eqk}-\pref{eqg} already in the static case, i.e. the system size $R\simeq 8$ mm and the Pearl length  $\L=2\l^2/d$, that at $T_{BKT}$  (where $\l^{-2}\simeq 0.5 \m$m$^{-2}$) is of the order of $\L\simeq 1$ mm. For a conventional value $D_e\simeq 10^{-5}m^2/s$ of the diffusion constant  $r_\o\simeq 1$ cm at 10 kHz, i.e. it is even larger than both $R$ and $\L$. Thus, the rounding effects at finite frequency on $\l^{-2}$ shown in Fig.\ \ref{fig-stand}c do not differ considerably from the ones found in the static case, so that the finite-frequency computation induces only a negligible shift of the transition temperature without accounting for the broad smearing of the jump observed in the experiments, see Fig.\ \pref{fig-stand}a.  Analogously finite-frequency effects lead now to a finite dissipation $\sigma_1$, but the $\d^{-2}$ response appears as almost a delta-like peak at $T_{BKT}$, in contrast to the wide dissipation signal observed in the experiments, see Fig.\ \pref{fig-stand}b. 

The failure of the standard BKT dynamical theory for a homogeneous system shown in Fig.\ \ref{fig-stand}
is a clear indication that some crucial ingredient is missing. It is worth noting that the same theoretical approach was shown to be instead in very good quantitative agreement with experimental data in
He films investigated in the past.\cite{ambegaokar_prb80,bishop_prb80}
One crucial difference between superfluid films and superconducting ones is
that in the latter case vortex-antivortex interactions are screened out by
charged supercurrenst, so that BKT physics becomes visible only for thin
enough films.\cite{review_minnaghen,review13} However, as the film thickness is reduced also the disorder
level increases, putting thin BKT films on the verge of the SIT, where
additional physical effects emerge. The most important one is the natural
tendency of the system to form inhomogeneous SC structures, that modify
crucially the above results derived for a purely homogeneous
superconductor. In particular, as it has been discussed in a series of
recent publications,\cite{mondal_prl11,yong_prb13} the inhomogeneity of the
system is the main reason for the smearing out of the universal superfluid jump with
respect to the BKT prediction (see Fig.\ \ref{fig-stand}a). However, as we shall see in the next Section, the analysis of the dissipative part shows that the inhomogeneity can have also a strong effect  on the ability of vortices to diffuse under an ac field, explaining the anomalously large resistive signal observed in the experiments.

\section{Dynamical BKT theory in the presence of inhomogeneity}

\begin{figure}[htp]
\includegraphics[scale=0.3,angle=0,clip=]{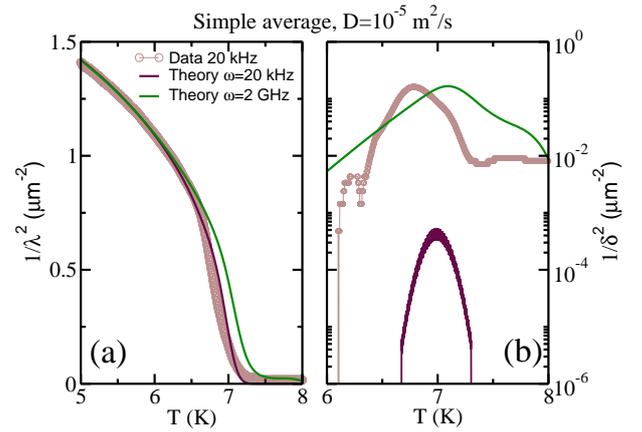}
\caption{(Color online) Comparison between the experimental data and the
  complex conductivity computed as the average over an inhomogeneous
  Gaussian distribution ($\s_G/\bar J=0.015$) of local superfluid-stiffness
  values for a conventional value of $D=10^{-5}$ m$^2$/s. While this procedure works very well for the inductive response
  (a), leading to the smearing of the BKT jump, it does not capture at all
  the intensity of the dissipative response (b) observed experimentally in
  the kHz frequency range.}
\label{fig-average}
\end{figure}

A first root to account for inhomogeneity is a direct extension of the procedure proposed in Refs.\
[\onlinecite{mondal_prl11,yong_prb13}] in the static case, i.e. an average of the conductivity over a distribution $P(J_i)$ of local superfluid-stiffness values $J_i$, that mimic in a simple
macroscopic picture the spatial inhomogeneity observed by STM.\cite{sacepe_11,pratap_13,noat_prb13} Here we extend this approach to the complex conductivity $\s(\o)$, computed for each patch according to Eq.\ \pref{defs}, while the global one is given by an average $\sigma=\int dJ_i P(J_i) \sigma_i(\o)$ over a Gaussian distribution $P(J_i)=(1/\sqrt{2\pi}\s_G)\exp[-(J_i-\bar J)^2/2\s_G^2]$ of local superfluid-density values. Here each $J_i(T)$ follows from the numerical solution of the BKT RG equations, using as initial value a BCS expression, such that $\bar J(T)$ has as starting point the BCS fit of low-temperature data shown in Fig.\ \ref{fig-expt2} (see details in Ref. [\onlinecite{yong_prb13}]). The complex conductance of each patch follows then from Eq.\ \pref{defs} above. The resulting $\s(\o)$ for a conventional value of the diffusion constant, i.e. $D\sim 10^{-5}$ m$^2$/s, is  shown inFig.\ \ref{fig-average}a. As we discussed above, finite-frequency effects have a negligible impact on the stiffness $\l_i^{-2}$ of each patch (see Fig.\ \ref{fig-stand}c), so at $\o=$20 kHz the average superfluid response is practically the same computed in Refs. [\onlinecite{mondal_prl11,yong_prb13}] for the static case. As one can see, since within patches with smaller/larger local $J_i$ values the transition occurs at lower/higher temperatures, the average stiffness displays a smeared transition, in good agreement with the experimental data. This support the notion, already pointed out in previous work,\cite{mondal_prl11,yong_prb13} that the main source of the superfluid-density jump smearing is the inhomogeneous distribution of local transition temperatures, while finite-frequency effects are not so relevant.
When applied to the dissipative finite-frequency vortex response the averaging procedure has the analogous effect: the delta-like peak found at $20$ kHz for $\d_i^{-2}$ in each patch is now convoluted with a Gaussian distribution of local transition temperatures, leading to a widening of the peak. However, this has also the unavoidable effect of reducing the peak intensity, that is now at $\o=20$ kHz two orders of magnitude {\em smaller} than what observed experimentally, see Fig.\ \ref{fig-average}b. In addition, while the inductive response is weakly dependent on the frequency, the dissipative one increases strongly when we move in the microwave regime, even after the average procedure. The peak in $\delta^{-2}$ shown in Fig.\ \pref{fig-average}b at $\o=2$ GHz for a conventional value of the vortex diffusion constant is instead of the correct order of magnitude, even if the shape of the peak is still different.

The strong disagreement between the experiments and the theory in the kHz frequency range suggests that the length scale $r_\o$ for vortex diffusion is the same expected in the microwave regime, so that $D$ must be taken in our simulation a function of frequency strongly decreasing at low $\o$. Notice also that the simple average procedure leads to an overestimate of the transition width in $\sigma_1$ at large frequency, see Fig.\ \pref{fig-average}b. To improve the treatment of the finite-frequency effect we then resort to a self-consistent effective-medium
approximation\cite{ema} (SEMA)  for the optical conductivity of the
inhomogeneous system. Thus, once generated a distribution of local
$\sigma_i(\o)$ conductance with probability $P_i\equiv P(J_i)$ the complex
SEMA conductance $\s(\o)$ is computed as the solution of the following equation:
\be
\lb{sema}
\sum_i P_i \frac{\s_i(\o)-\s(\o)}{\s_i(\o)+\eta\s(\o)}.
\ee
where $\eta=D-1$ coincides with 1 in two spatial dimensions. 
Notice that Eq.\ \pref{sema} can also be rewritten as:
\be
\s=\left( (1+\eta)\sum_i \frac{P_i}{\s_i+\eta\s}\right)^{-1},
\ee
so that one sees that in the limiting case $\eta=0$ $\s$ is computed by
assuming that the complex impedances are in series, as it is the only
possible case in $D=1$ physical dimensions. In this situation the
contribution of each single resistor to the overall dissipative response in
enhanced: indeed, for $\eta=0$ at each temperature $\s$ is dominated by the
resistor $\s_i$ having a peak at that temperature, weighted as $1/P_i$
instead of $P_i$ that one had in the simple average procedure. Thus the use
of the effective-medium approximation amplifies in general the dissipative
response of our network of BKT complex conductances, getting a better
agreement with the experiments.  For what concerns instead $\s_2$ the SEMA
conductance behaves essentially as the averaged one, with a general
smoothening of the superfluid-density jump due to the superposition of
several $J_s^i(T)$ curves vanishing at different temperatures.

\begin{figure}[htp]
\includegraphics[scale=0.3,angle=-0,clip=]{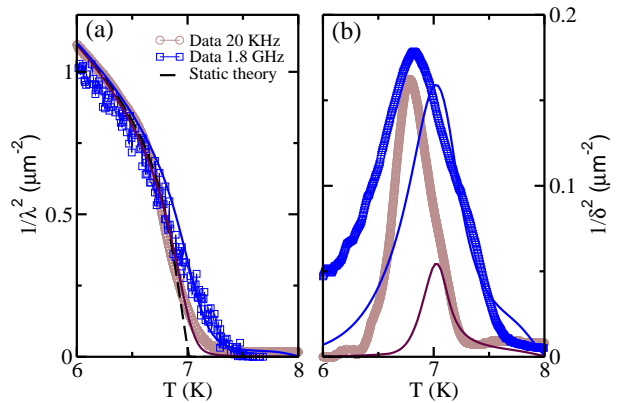}
\caption{(Color online) Inductive (a) and resistive (b) response of the
  SEMA complex conductivity from Eq.\ \pref{sema}. The same color is used
  for the experimental data (symbols) and the theoretical curves (solid
  lines). The dashed black line in panel (a) corresponds to the superfluid
  response in the static limit. The value of the diffusion coefficient $D$
  at the various frequencies is reported in the text. The remaining
  parameters for the BCS fit are the same of Fig.\ \ref{fig-expt2} and Fig.\ \ref{fig-stand}a, while
  the BKT parameters are slightly different due to the different averaging
  procedure, i.e. $\s_G/\bar J=0.02$ and $\mu/\bar J=1.18$. }
\label{fig-sema}
\end{figure}

In Fig.\ \ref{fig-sema} we show results for the complex SEMA conductance
obtained with $D\simeq 10^{-11}$ m$^2$/s in the kHz regime, corresponding
to a $r_\o\simeq 8\xi_0$ at $\o=20$ kHz. In the kHz regime the dissipative
response of each local impedance is highly enhanced with respect to the
results of Fig.\ \ref{fig-average}, and the $\s_1$ peak is in better
agreement with the experiments. In the microwave regime instead $D$
increases, and a good compromise between the fit of the inductive and
dissipative response is found for a conventional value $D=1.5\times 10^{-5}$
m$^2$/s, corresponding to $r_\o\simeq \xi_0$ at $\o=1.8$ GHz.  Observe
however that still the experimental data of Fig.\ \ref{fig-expt} show a dependence on
the probing frequency more pronounced than what found theoretically.
This is a consequence of the weak (logarithmic) dependence of the vortex
dielectric function on the scale $r_\o$, as evidence for example by the
estimate \pref{dtbar} of the peak temperature of each resistor.

\section{Discussion and Conclusions}

As one can see in Fig.\ \ref{fig-sema}, even accounting for the slow vortex
diffusivity and the system inhomogenity the agreement with the experimental
results for $\s_1$ in the kHz regime is not as good as the one for $\s_2$. On the other hand,
the present phenomenological analysis, where the spatial inhomogeneity of
the system is included in an effective-medium approach with a
frequency-induced length-scale cut-off, elucidates already the emergence of
a peculiar interplay between the inductive and resistive response near the
BKT transition. In particular, by having in mind also theoretical results on disordered films.\cite{seibold_prl12} our results can be interpreted by assuming
that the superfluid inductive response can take advantage of the existence
of preferential (quasi one-dimensional) percolative paths connecting the
good SC regions, while thermally-excited vortices responsible for the
dissipation will mainly proliferate far away from the SC region. This
notion is further supported by the emergence of an intrinsic length scale
for vortex diffusion $r_\o\sim \xi_V\sim 10-50$ nm that correlates very
well with the typical size of the SC granularity observed experimentally by
STM experiments,\cite{sacepe_11,pratap_13,noat_prb13} and predicted
theoretically in microscopic models for
disorder.\cite{ioffe,nandini_natphys11,seibold_prl12,lemarie_prb13} Such a
mechanism could thus explain the coexistence of both a large superfluid and
dissipative response in a wide temperature range around the transition.

The quantitative comparison presented here between the theoretical prediction of a conventional BKT approach for homogeneous films and the experiments shows also that some care must be taken while analysing the data with a standard scaling approach,\cite{ffh} as suggested for example for microwave measurements in thin InOx films in Ref.\ [\onlinecite{armitage_prb11}]. Indeed, we have shown that the shape of the complex conductivity is strongly affected by the inhomogeneous distribution of local SC properties. For example, even in the microwave regime the broadening of the superfluid-density jump cannot be understood as a trivial finite-frequency effect, as it is instead assumed in the usual scaling hypothesis.\cite{ffh,armitage_prb11} Thus, the extraction\cite{armitage_prb11} of a scaling frequency $\o_0(T)$ that correlates with the usual BKT behaviour \pref{xi} of the vortex correlation length does not mean in general that a standard BKT scaling, i.e. the one predicted in the homogeneous BKT theory, is at play. Indeed, the real agreement with BKT scaling should be proven by comparing in a qualitative and quantitative way the complex conductance itself. In our case, we checked explicitly that even though the microwave data can be rescaled to give a BKT-like scaling frequency $\o_0(T)$, the scaling function itself deviates strongly from the BKT one, since its shape is controlled by the inhomogeneity. In addition, while we focused here on thin film where the transition has BKT character, the peculiar role of inhomogeneity is a general feature of disordered films near the SIT. Indeed, the strong slowing down of the fluctuation conductivity observed recently in thick NbN films,\cite{mondal_scire13} where BKT physics is absent, show that the enhanced finite-frequency effects reported near the SIT can be analogously interpreted as a signature of an intrinsic length-scale dependence associated to inhomogeneous SC domains. 

Finally, the present analysis clarifies also that the absence of a sharp superfluid-density jump in thin films of conventional superconductors cannot be attributed\cite{yazdani_prl13} to the the mixing between inductive and reactive response that already occurs at the kHz frequency. Indeed, in the standard homogeneous case the effect of the frequency on the universal jump would be the one shown in Fig.\ \ref{fig-stand}c, i.e. the $\l^{-2}$ should still drop to zero so rapidly to appear as a discontinuous jump in the experiments. The rounding effect of the stiffness are instead entirely  due to the inhomogeneity, that also mixes in a non-trivial way inductive and reactive response, making it meaningless the extraction from the data of the inverse inductance in order to analyse the superfluid-density jump, as done e.g. in Ref. [\onlinecite{yazdani_prl13}] (see also Appendix A).

In summary, we analysed the occurrence of the dynamical BKT transition in thin films of NbN. We measured the same samples both in the kHz regime, where dynamical effects should be negligible according to the standard view, and in the microwave regime, where one would expect instead a sizeable finite-frequency induced dissipative response. Our experimental results show a consistent broadening of the universal BKT superfluid-density jump, that can be attributed to inhomogeneity, and an anomalously large resistive response in the low-frequency range, that cannot be understood by means of a standard value of the vortex diffusion constant. By making a quantitative comparison between the experiments and the theoretical predictions for the BKT physics in a inhomogeneous SC environment we show that the dissipative response in the kHz regime can only be understood by assuming a low vortex diffusivity. This effect limits the vortex motion over an intrinsic length scale of the order of the typical size of homogeneous SC domains observed by STM near the SIT. While the present approach accounts for the emergence of SC inhomogeneity in a phenomenological way, a more microscopic approach is certainly required to understand how the BKT vortex physics can accommodate to the disorder-induced SC granularity by preserving its general character. 

\acknowledgments
We acknowledge P. Armitage and C. Castellani for useful discussions and suggestions and J. Jesudasan for his help in sample preparation. 
D.C. thanks the Visiting Students Research Program for hosting his visit in TIFR, Mumbai during the course of his work.
L.B. acknowledges financial support by MIUR under projects FIRB-HybridNanoDev-RBFR1236VV, PRIN-RIDEIRON-2012X3YFZ2 and Premiali-2012 ABNANOTECH.

\appendix
\section{Dielectric function of vortices within the RG approach}

The expression for the vortex dielectric constant $\e(\o)$ that appears in Eq.\ \pref{defs} has been derived\cite{ambegaokar_prb80,HN}
by exploiting the analogy between the Coulomb gas and the vortices. It contains two contribution, one $\e_b$ due to bound vortex-antivortex pairs that exist already below $T_{BKT}$, and one $\e_f$ due to free single-vortex excitations that are thermally excited above $T_{BKT}$. To compute $\e_b$ one exploits the idea that under the applied oscillating field vortices experience a Langevin dynamics controlled by the diffusion constant $D$. In practice, if we think that in the BKT theory what controls $J_s(r)$ is the screening due to neutral vortex-antivortex pairs at a scale $r$, dynamics introduces one additional scale $r_\o$ such that pairs with separation $r\gg r_\o$ will not contribute to the polarization since they will change the relative orientation over a cycle of the oscillating field. This explains while $\e_b$ is cut-off at $r_\o$. On the other hand free vortices have uncorrelated motions with respect to each other and, when present, they will contribute directly to dissipation. The general expressions for the bound $\e_b$ and free $\e_f$ vortex contributions are then:\cite{ambegaokar_prb80}
\bea
\lb{epsb}
\e_b(\o)&=&1+\int_{\xi_0}^\xi dr \frac{d\tilde\e(r)}{dr} \frac{14 D
  r^{-2}}{-i\o+14 D r^{-2}}\\
\lb{epsf} 
\e_f(\o)&=&i \frac{4\pi^2 J}{k_B T}\frac{D n_f}{\o}=i\frac{2\pi^2 J}{7k_BT}\left(\frac{r_\o}{\xi}\right)^2
\eea
where  $n_f$ is the free-vortex density, expressed in terms of the correlation length $\xi$ as $n_f={1}/{\xi^2}$, $J$ is the superfluid stiffness defined in Eq.\ \pref{defj} above and $\tilde\e$ is defined in terms of the RG variable $K$ \pref{eqk} as
\be
\lb{tildeeps}
\tilde\e(r)=\left.\frac{K(0)}{K(\ell)}\right|_{\ell=\ln(r/\xi_0)}.
\ee

Let us first analyze the case of an infinite system at $\o=0$. When the
system is infinite the correlation length $\xi=\infty$ below $T_{BKT}$. In this case the upper limit of
integration in Eq.\ \pref{epsb} is set at $\infty$ and the free-vortex
contribution is different from zero only above $T_{BKT}$. Morevoer, by
using $\o=0$ $\e_b$ (and then also the total $\e$) is
purely real and it can be easily computed using the definition
\pref{tildeeps}: indeed we have that 
\bea
\e_b(\o=0)&=&1+\int_{a_0}^\infty dr
\frac{d\tilde\e(r)}{dr}=1+\tilde\e(\infty)-\tilde\e(\xi_0)\nn\\
\lb{epsb0}
&=&\tilde\e(\infty)=\frac{K(0)}{K(\infty)}=\frac{J(0)}{J_s}
\eea
in agreement with Eq.\ \pref{eps0} above, where we already used the definition \pref{defjs} $J(\ell=\infty)=J_s$. 
Since $J(0)\propto n_s^0$ while $J_s\propto n_s$, where $n_s$ is the real superfluid density including also the vortex contribution we obtain in  Eq.\ \pref{defs} that at $\o=0$ the response is purely inductive:
\be
\s(\o=0)=-\frac{n_s d e^2}{i\o m}.
\ee

At finite frequency the dielectric function develops an imaginary
part, responsible for the dissipative response detected via
$\sigma_1$. Bound vortices give the main contribution to the real part
of the dielectric function, while free vortices occurring on the
length scale $r_\o$ contribute to the imaginary part of the dielectric
function. All the theoretical results shown in the mansuscript have been obtained by means of the full numerical solution \pref{epsb}-\pref{epsf}, where $K(\ell)$ is the solution of the RG equations \pref{eqk}-\pref{eqg}. On the other hand, one can also provide a rough estimate of the expected behavior of complex conductivity based on the above formulas. For what
concern $\e_1$ the contribution of bound vortices at finite frequency can be estimated by replacing in Eq.\ \pref{epsb0} above the upper cut-off of integration with $r_\o$, that is that maximum distance explored by vortices under the applied field.\cite{ambegaokar_prb80} One then has:
\be
\lb{e1o}
\e_1(\o)={\e_b}_1(\o)\approx \frac{K(0)}{K(\ell_\o)}
\ee
so that instead of the discontinuous jump of $\e_1$ expected at $\o=0$ one
observes now a rapid but continuous downturn, as discussed in Sec. III. At
the same time for $\e_2$ one has the largest contribution from free
vortices, i.e. the contribution \pref{epsf} above that has been discussed
below Eq.\ \pref{e2app} in Sec. III. Notice that in the homogeneous case
Eq.\ \pref{defs} and \pref{e1o} above show that if one plot directly
the inverse inductance, as it has been sometimes
suggested,\cite{yazdani_prl13} i.e
 \be
 L^{-1}=\frac{n_s^0 e^2 d}{m\e_1(\o)}\approx \frac{n_s(r_\o) e^2 d}{m},
 \ee
 then one can directly access the superfluid-density jump occurring at finite frequency. However, while this would be a viable procedure to isolate the real and imaginary part of the dielectric function for a homogeneous system, it fails completely in the presence of inhomogeneity, that has a much more drastic effect on the jump than the finite-frequency behaviour. Thus in the case of thin disordered films the lack of a sharp BKT jump cannot be circumvented by extracting from the measured conductivity the inverse inductance, since this procedure mixes in an artificial and uncontrolled way the non-trivial effects of the inhomogeneity on the finite-frequency response.

\end{document}